\documentclass[11pt]{article}    
\usepackage{amsmath,amssymb,amsthm,amsfonts,natbib,color,url}
\usepackage{graphicx,url}
\usepackage{subcaption}
\usepackage{epstopdf}

\newcommand{\N}{{\mathbb N}}

\newcommand{\R}{\mathbb R}

\begin{document}


\title{Multi Loci Phylogenetic Analysis with Gene Tree Clustering}
\author{Kenji Fukumizu \and Ruriko Yoshida \and Chrysafis Vogiatzis}



\maketitle
\begin{abstract}
\textbf{Summary:} Both theory and empirical evidence indicate that phylogenies (trees) of different genes (loci) do not display precisely matched topologies. This phylogenetic incongruence is attributed to the reticulated evolutionary history of most species due to meiotic sexual recombination in eukaryotes, or horizontal transfers of genetic materials in prokaryotes. Nonetheless, most genes do display topologically related phylogenies; this implies they form cohesive subsets (clusters). In this work, we compare popular clustering methods, and show how the performance of the normalized cut framework is efficient and statistically accurate when obtaining clusters on the set of gene trees based on the geodesic distance between them over the Billera-Holmes-Vogtmann (BHV) tree space. We proceed to present a computational study on the performance of different clustering methods with and without preprocessing under different distance metrics and using a series of dimension reduction techniques.  \\
\textbf{Results:} First, we show using simulated data that indeed the Ncut framework accurately clusters the set of gene trees given a species tree under the coalescent process. We then depict the success of our framework by comparing its performance to other clustering techniques, including k-means and hierarchical clustering. The main computational results can be summarized to the stellar performance of the Ncut framework even without dimension reduction, the similar performance portrayed by k-means and Ncut under most dimension reduction schemes, the utter failure of hierarchical clustering to accurately capture clusters, as well as the significantly better performance of the NJp method, as compared to MLE. \\
\end{abstract}

\section{Introduction}

During especially this last decade, the field of phylogenetics has been undergoing a gradual paradigm shift away from the notion of the strictly bifurcating, completely resolved species trees, to a recognition that species are containers of allelic variation for each gene. It is very well established that differences in lineage sorting due to genetic drift lead to differences in phylogenetic tree topologies \citep{Maddison1997}. Gene flow in ancestral populations and independent lineage sorting of polymorphisms is fully expected to generate topological conflicts between gene trees in reticulating (e.g., sexually recombining) species \citep{HusonSteel, Weisrock, Taylor2000}. Both extant and ancestral species could exhibit this phenomenon, so ancestral species should not be regarded as node points in a fully resolved bifurcating tree, but instead can be thought of as spatiotemporal clouds of individual genotypes with all their inherent allelism. Thus, a central issue in systematic biology is the reconstruction of populations and species from numerous gene trees with varying levels of discordance \citep{Brito2009,Edwards2009}. While there has been a well-established understanding of the discordant phylogenetic relationships that can exist among independent gene trees drawn from a common species tree \citep{Pamilo1988,Takahata1989,Maddison1997,Bollback2009}, phylogenetic studies have only recently begun to shift away from single gene or concatenated gene estimates of phylogeny towards these multi-locus approaches (e.g. \citet{Carling2008,Yu2011b,Betancur2013,Heled2013,Thompson2013}).


There exist numerous processes that can reduce the correlation among gene trees. Negative or balancing selection on a particular locus is expected to increase the probability that ancestral gene copies are maintained through speciation events \citep{Takahata}. Horizontal transfer shuffles divergent genes among different species \citep{Maddison1997}. Correlation may also be reduced by naive sampling of loci for analysis. For example, paralogous gene copies will result in a gene tree that conflates gene duplication with speciation. Similarly, sampled sequence data that span one or more recombination events will yield ``gene trees'' that are hybrids of two or more genealogical histories \citep{PosadaCrandall2002}. These non-coalescent processes can strongly influence phylogenetic inference \citep{PosadaCrandall2002, MartinBurg2002, Edwards2009}. In addition, \cite{Rivera} showed that an analysis of complete genomes indicated a massive prokaryotic gene transfer (or transfers) preceding the formation of the eukaryotic cell, arguing that there is significant genomic evidence for more than one distinct class of genes. These examples suggest that the distribution of eukaryotic gene trees may be more accurately modeled as a mixture of a number of more fundamental distributions. In order to find a mixture structure in distributions of gene trees, we first need to find how many components of distributions there are in the mixture. This is the main reason why in this work we focus on the problem of clustering gene trees over the ``tree space''.

Many researchers take an approach to apply a likelihood based method, such as the \emph{maximum likelihood estimator} (MLE) or \emph{Bayesian inference} on the {\em concatenated alignment} from gene alignments in order to reconstruct the species tree.  However, \citet{Roch2014} showed that if we apply a likelihood based method on the concatenated alignment from gene alignments, then the resulting trees might be statistically inconsistent because some gene trees are significantly incongruent from the species tree due to incomplete lineage sorting, horizontal gene transfer, among other reasons.  More precisely, they showed that under the multi-species coalescent with a standard site substitution model, such as the general time reversible (GTR) model  \cite[]{Tavare1986} etc, the MLE on a sequence data concatenated across genes under the assumption that all sites have evolved independently and identically on a fixed tree  is a statistically inconsistent estimator of the species tree.


Typically, statistical analysis on phylogenetic trees is conducted by mapping each tree to a vector in $\R^d$, $d\in \N$: this is referred to as a {\em dissimilarity map}. Given any tree $T$ of $n$ leaves with branch length information, one may produce a corresponding {\em distance matrix}, $D(T)$. This distance matrix is an $n\times n$ symmetric matrix of non-negative real numbers, with elements corresponding to the sum of the branch lengths between pairs of leaves in the tree. To calculate $D_{(ij)}(T)$, one simply determines which edges of the tree form the path from a leaf $i$ to a leaf $j$, and then sums the lengths of these branches. Since $D(T)$ is symmetric and has zeros on the diagonal, the upper-triangular portion of the matrix contains all of the unique information found in the matrix. We can vectorize a tree $T$ by enumerating this unique portion of the distance matrix, \[v_D(T) := (D_{12}(T), D_{13}(T), \ldots , D_{23}(T),\ldots,D_{n-1, n}(T)) \] which is called the {\em dissimilarity map} of a tree $T$ and is a vector in $\R^{{n \choose 2}}$.

However, the space of phylogenetic trees with $n$ leaves is not a Euclidean space. In fact, it is represented as the union of lower dimensional polyhedral cones in $\R^{{n \choose 2}}$. \citet{Billera2001} introduced a continuous space which explicitly models the set of rooted phylogenetic trees with edge lengths on a fixed set of leaves.  Although the Billera-Holmes-Vogtmann (BHV) tree space is not Euclidean, it is non-positively curved, and thus any two points are connected by a unique shortest path through the space, called a {\em geodesic}. In this computational study, we show, among other things, that using the BHV tree space can help produce more statistically accurate results. 

This paper focuses on such tree spaces and presents a computational study on multi-loci phylogenetic analysis using gene tree clustering over the BHV tree space. More specifically, this work presents the differences and the performance of a normalized cut framework, $k$-means, and hierarchical clustering in the Euclidean and the BHV tree space, and under different dimension reduction approaches using simulated datasets. The paper is organized as follows. Section \ref{sec:fundamentals} offers a basic review of the BHV space, the normalized cut framework, and the dimension reduction for the interested reader. In Section \ref{results}, we present our computational study and the results we obtained using simulated datasets. Moreover, Section \ref{discussion} discusses the results and focuses on our main computational results, while also summarizing our work and offering insight for possible future directions.

\section{Fundamentals}
\label{sec:fundamentals}

In this paper, we present a comparative study of different methods for multi-loci phylogenetic analysis using gene tree clustering based on the distance matrix obtained by the geodesic distances between two trees over the BHV space. The methods we compare are the normalized cut framework, based on the seminal contribution by \citet{ShiMalik2000}, $k$-means (e.g., \cite{schenker2003classification}), and hierarchical clustering (the interested reader is referred to \cite{maimon2005data} for an excellent overview). 

Furthermore, we investigate how dimension reduction methods can be applied in order to extract a lower dimensional structure before clustering and whether that affects the quality of solutions compared to applying our clustering methods directly upon the original distance matrix. This reduction also helps with the visualization of the data. For dimension reduction, kernel principal component analysis (KPCA \citet{KernelPCA}) and $t$-stochastic neighborhood embedding (t-SNE \citet{vanderMaatenHinton2008_tsne}) are employed among many other methods. Hereafter, we refer to the above three branches as {\em direct, KPCA}, and {\em t-SNE}, respectively.  

We now proceed to offer some basics on the BHV tree space, the normalized cut problem, and the different dimension reduction techniques used herein.

\subsection{Billera-Holmes-Vogtmann Tree Space}
\cite{Billera2001} introduced a continuous space that models
the set of rooted phylogenetic trees with edge lengths on a fixed set
of leaves. (Unrooted trees can be accommodated by using either the Ferras
transform, or by designating an arbitrary leaf node as the root.)
It is known that in the {\em Billera-Holmes-Vogtmann (BHV) tree space} any two points
are connected by a geodesic, and the distance between two trees is defined as the
{\em length of the geodesic} connecting them.


Consider a rooted tree with $n$ leaves. Such a tree has at most $2n -2$ edges;
there are $n$ terminal edges, which are connected to leaves,
and as many as $n-2$ internal edges. The maximum number of edges is
achieved when the tree is binary, but the number of edges can be lower if the tree
contains any polytomies. With each distinct tree topology, we
associate a Euclidean {\em orthant} of dimension equal to the number
of edges that the topology possesses. (Here, we may regard an orthant
to be the subset of $\R^d$ with all coordinates non-negative.) For each
topology, the orthant coordinates correspond to edge lengths in the
tree.

Since all tree topologies have the same set of $n$ terminal leaves,
and each of these leaves is associated with a single terminal edge,
the orthant coordinates associated with the terminal edges are of
less interest than those of internal nodes. As a result, we will
simplify our discussion by ignoring the terminal edge lengths, and
concern ourselves primarily with the portion of each orthant which
describes the internal edges. (Recall that this space has at most
$n-2$ dimensions.)

Since each of the coordinates in a simplified orthant corresponds to
an internal edge length, the orthant boundaries (where at least one
coordinate is zero)  represent trees with collapsed internal
edges. These points can be thought of as trees with slightly
different---but closely related---topologies. The BHV
space is constructed by noting that the boundary trees from two different
orthants may describe the same polytomic topology. With this
insight, we may set about constructing the space by grafting
orthant boundaries together when the trees they represent coincide.


Since each orthant is locally a Euclidean space, the shortest path
between two points within a single orthant is a straight line. The
difficulty comes in establishing which sequence of orthants joining
the two topologies will contain the geodesic. In the case of four
leaves, we could do this through a brute-force search, but we cannot
hope to do so with larger trees. \cite{owen2011fast} present a
quartic-time algorithm (in the number of leaves $n$) for finding the
geodesic path between any two points in the space.  Once
the geodesic is known, computing its length---and thus the distance
between the trees---is a simple matter.

\subsection{Clustering}

Given a set of gene trees for the species in analysis, a clustering algorithm is applied based on the distance matrix containing the geodesic distances in the BHV tree space. As an alternative, dimension reduction may be applied before the clustering when directly applying the clustering techniques proves unfruitful. This is also helpful for visualization purposes.  For
the details of the clustering and dimension reduction methods, see the
Supplementary Material.

There are many standard clustering methods raging from non-hierarchical clustering such as k-means to hierarchical clustering methods.  Among others, this paper considers three methods: normalized cut, k-means, and hierarchical clustering (average linkage).   The k-means is the most standard non-hierarchical clustering method, which has been used in a large number of applications.  Note, however, that with BHV geodesic distance the k-mean methods is not computationally feasible, because update of centroids required in the methods is too expensive.  As a linkage method for hierarchical clustering we use average linkage, since it is applied to general distance or dissimilarity measures.  

From the clustering methods presented in this computational study, the normalized cut framework \citep{ShiMalik2000} is applied for the first time in phylogenetics, to the best of our knowledge. Normalized cut can be employed for clustering using only a similarity or dissimilarity matrix; we can observe that the coordinates of the original data points are not necessary. To properly apply the normalized cut framework in a clustering setting the only required input is the set of data points (each represented by a node in an undirected graph) forming node set $V$, and a set of weights of similarity between them (the edge set, $E$, of the graph). Then, the normalized cut framework aims to detect a bipartition of the node set of the graph in two node sets, $(S, \bar S)$, such that \eqref{NCutCriterion} is minimized with $S\cup \bar S=V$ and $S\cap\bar S=\emptyset$.

\begin{align}
\label{NCutCriterion}
 NCut(S,\bar{S}) = \frac{cut(S, \bar{S})}{assoc(S,V)} + \frac{cut(S, \bar{S})}{assoc(\bar{S},V)}
\end{align}

More recently, the problem has been studied by \cite{hochbaum2010polynomial} and \cite{hochbaum2013polynomial}, where normalized cut variants are discussed, with some of them being shown to be solvable in polynomial time. Among them, of interest to the clustering community would be the \emph{``normalized cut"} problem of \cite{sharon2006hierarchy}, which is nothing more but a single version of the original normalized cut criterion shown in \eqref{NCutCriterion} and the \emph{ratio regions} problem \citep{cox1996ratio}. In \cite{hochbaum2010polynomial} both the ratio regions and ``normalized cut" problems were shown to be poly-time solvable. The normalized cut framework has been successfully applied to numerous applications, including image segmentation \citep{ShiMalik2000,Carballido-Gamio_etal2004,Yao_etal2012}, biology \citep{XingKarp2001,Higham_etal2007}, and social networks \citep{Newman2013PhysRevE}.

The normalized cut is known to be solved approximately (with typically good performance) as a generalized eigenproblem, which admits a straightforward and easy to implement solution. In our experiments, for simplicity, the similarity $w_{ij}$ is given by $w_{ij}=\exp(-\frac{1}{2\sigma^2}D_{ij}^2)$, where $D_{ij}$ is the distance matrix, and $\sigma=1.2\times \text{Median}\{D_{ij}\mid i\neq j\}$.

\subsection{Dimension reduction}

As an optional procedure before clustering, a low dimensional expression of gene trees may be extracted from the distance matrix.  Among various dimension reduction methods, kernel principal component analysis \cite[KPCA, ][]{KernelPCA} and t-stochastic neighborhood embedding \citep[t-SNE, ][]{vanderMaatenHinton2008_tsne} are chosen for our analysis by preliminary experiments (we applied also spectral methods and Isomap, but the results were less favorable than KPCA and t-SNE.) Those methods extracted three dimensional expression of data, when applied, and the Ncut was applied to the Euclidean distance matrix among the three dimensional data points.

KPCA is a nonlinear extension of the standard principal component analysis (PCA); it applies PCA to feature vectors, which are given by nonlinear mapping of the original data to a feature space.  The nonlinear map is defined by a {\em positive definite kernel}, and the feature space is a possibly infinite dimensional Hilbert space provided implicitly by the positive definite kernel.  KPCA gives nonlinear functions $f_1,\ldots f_d$ of data points $(X_i)_{i=1}^N$ so that $(f_1(X_i),\ldots,f_d(X_i))_{i=1}^N$ can serve as a $d$-dimensional representation of data.   The analysis of this paper uses Gaussian kernel $k(X_i,X_j)=\exp(-\frac{1}{2\sigma^2} D_{ij}^2)$ where $D_{ij}$ is the distance matrix of the gene trees\footnote{While this kernel with an arbitrary distance matrix $D$ is not necessarily positive definite, in our analysis the Gram matrices $k(X_i,X_j)$ made by the given data were positive definite.}.

t-SNE is a method for low-dimensional expression or visualization of high-dimensional data; it typically extracts two or three dimensional expression.  Given $(X_i)_{i=1}^N$ in a high-dimensional space, t-SNE first computes a probability $p_{ij}$ based on the distance matrix so that a high probability implies similarity of $X_i$ and $X_j$.  The method then provides a low dimensional expression $(Y_i)_{i=1}^N$ in such a way that a  probability $q_{ij}$ defined similarly for a pair $(Y_i,Y_j)$ is close to $(p_{ij})$.  The points $(Y_i)_{i=1}^N$ are found with numerical optimization to minimize the Kullback-Leibler divergence  between $(p_{ij})$ and $(q_{ij})$.
In our experiments, a Matlab implementation by van der Maaten ({lvdmaaten.github.io/tsne/}) is used.  The perplexity parameter, which gives a way of determining local bandwidth parameters, is set 30 in our experiments.

\section{Results}
\label{results}
We conducted numerical experiments with simulated datasets and a genome dataset. 
All our simulated datasets were generated using the software {\tt Mesquite} \cite[]{Maddison:2009wa}. We first demonstrate that the normalized cut framework accurately clusters the set of gene trees given by a species tree under the coalescent process. Then, we proceed to compare different dimension reduction schemes and their performance as compared to clustering via normalized cut directly on the original tree space. Last, we compare $k$-means and hierarchical clustering to our proposed approach. Two main observations throughout our obtained results are that hierarchical clustering is not effective in recognizing clusters (as opposed to normalized cut and $k$-means), and that the frameworks perform better on the gene trees reconstructed via the neighbor-joining (NJ) method \cite[]{Saitou1987} than those reconstructed via the MLE under evolutionary models. 

The experimental design for the genome dataset in Subsection 3.2 is as follows. The three clustering methods were first applied to a genome-wide dataset on coelacanths, lungfishes, and tetrapods from \citet{Amemiya,Liang2013}, where it was observed that there were two reliable clusters in their 1290 genes. Based on the datasets, we reconstructed the consensus trees using NJ trees with bootstrap confidence for the clusters $\geq 0.95$ (see Subsection \ref{subsec:lungfish} for more details). We performed numerical experiments using the (typically used) Euclidean space and the (better suited) BHV tree space and compared the accuracy of the two spaces for the goal of recognizing more statistically accurate clusters. Last, we compared different preprocessing dimension reduction schemes for each and every one of the spaces, and the clustering techniques. 

Overall, we obtained consistent results with both the normalized cut and the $k$-means frameworks on the consensus trees obtained. The consensus tree from one cluster (of 858 gene trees with the direct application of the normalized cut, of 761 gene trees with the normalized cut after applying KPCA, and of 817 gene trees with t-NSE normalized cut) supports the view of \citet{Fritzsch,Gorr} that claims that coelacanths are most closely related to the tetrapods; furthermore, the consensus tree constructed from the other cluster (of 322 gene trees with the direct Ncut algorithm, of 320 gene trees with the KPCA Ncut algorithm, and of 463 gene trees with the t-NSE Ncut) supports the view of \citet{Takezaki}, that is, the coelacanth, lungfish, and tetrapod lineages diverged within a very short time interval and that their relationships may represent an irresolvable trichotomy. We now proceed to describe our results in more detail on both simulated datasets (see Subsection \ref{sec:results_simul}) and coelacanths, lungfishes, and tetrapods (see Subsection \ref{subsec:lungfish}).

\subsection{Simulated data sets}
\label{sec:results_simul}

The simulated data is generated as follows.  We have fixed the population size $N_e = 10,000$ and we set the species depth $c \cdot N_e$ where $c = 0.6, 0.7, 0.8, 0.9, 1, 1.2, 1.4, 1.6, 1.8, 2$.  Then for each species depth $c \cdot N_e$, we generated 100 species trees from the Yule process and we picked randomly two trees from them.  With each species tree, we generated 1000 random gene trees under the coalescent process within the species tree using the software {\tt Mesquite} \cite[]{Maddison:2009wa}.  To generate the sequences we have used the software {\tt PAML} \cite[]{Yang1997} under the Jukes-Cantor (JC) \cite[]{Jukes1969} $+\Gamma$ model with $\kappa = 4.0$, and the number of categories of the discrete gamma model is 1 with $\alpha = 1.0$ for one set of gene trees under the first species tree and we generated sequences under the GTR  $+\Gamma$ model with $\kappa = 4.0$, and the number of categories of the discrete gamma model is 1 with $\alpha = 1.0$ for the other set of gene trees under the second species tree.  The frequencies for T, C, A, and G in the data are set as $0.15, \, 0.35, \, 0.15, \, 0.35$, respectively.  We set the length of sequences as 500. To reconstruct trees from these DNA sequences, we used the NJ algorithm with the p-distance \cite[]{Saitou1987} (we call NJp method from here on) to reconstruct the NJ trees, and used the software {\tt PHYML} \cite[]{phyml} to reconstruct MLE trees under the GTR model \cite[]{Felsenstein1981}, the Hasegawa-Kishino-Yano (HKY) \cite[]{Hasegawa1985} model, and the Kimura 2 parameter (K80) model \cite[]{Kimura1980}; they are denoted by MLE-GTR, MLE-HKY, and MLE-K80, respectively, from now on.

\begin{figure}[t]
\centering
  \includegraphics[scale=0.25]{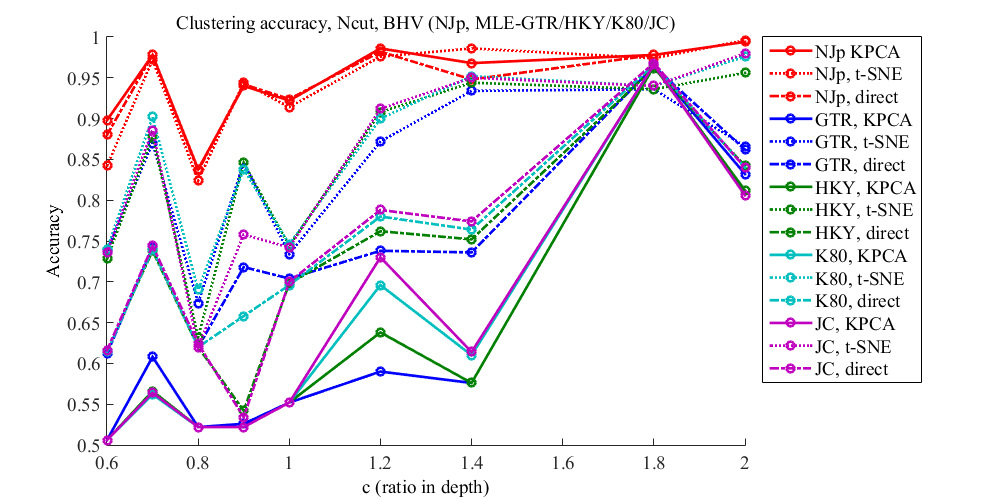}\vskip 3mm
  \caption{Ncut Clustering accuracy for simulated data. NJp gives superior accuracy than MLE. The results of MLE show three groups depending on the three clustering methods.}\label{fig:cluster_acc}
\end{figure}

Fig.~\ref{fig:cluster_acc} shows the rates of correctly clustered genes by the three newly proposed clustering schemes: direct Ncut, KPCA Ncut, and t-SNE Ncut. Here and in the sequel, {\it direct} means direct application of a clustering methods without dimension reduction.  Generally the accuracy is higher for larger species depths, which imply clearer separation.  There is a significant difference of accuracy between NJp and MLE tree reconstruction methods; the NJp method (solid lines) gives better clustering for all the three clustering methods.  It is also noted that the accuracy for the MLEs has clear groups based on the clustering schemes; t-SNE Ncut (broken lines), direct Ncut (dashed lines), and KPCA Ncut (dotted lines) give groups of similar accuracy levels in this order.

\begin{figure}[t]
\centering
  \includegraphics[scale=0.25]{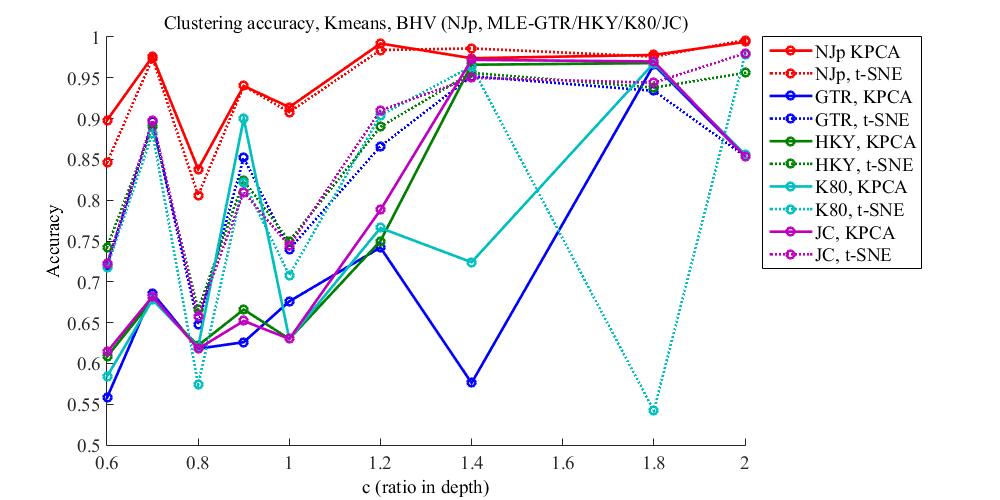}\vskip 3mm
  \caption{$k$-means Clustering accuracy for simulated data. It performs similarly to the normalized cut framework; the main difference is the lack of a direct application of $k$-means on the datasets. \label{fig:k-means}}
\end{figure}

\begin{figure}[t]
\centering
  \includegraphics[scale=0.25]{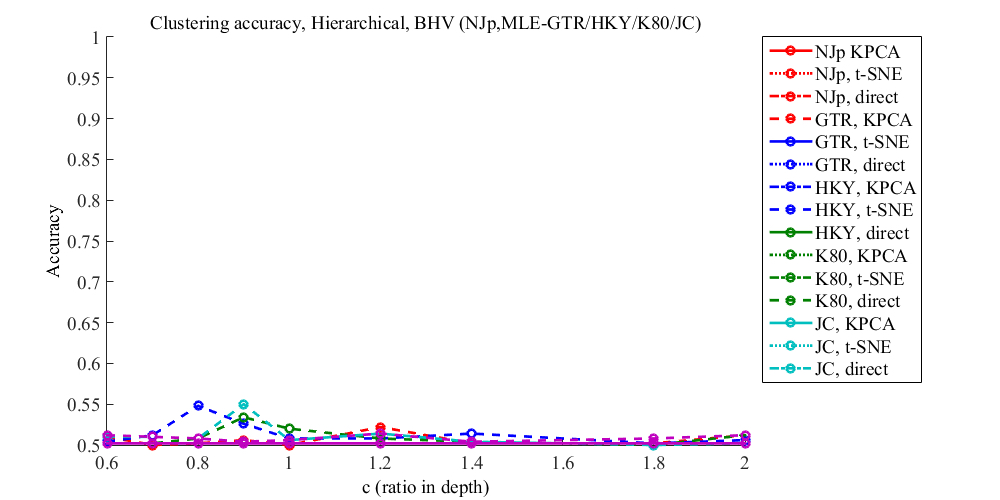}\vskip 3mm
  \caption{Hierarchical Clustering accuracy for simulated data. It is easy to see that the results obtained are statistically far worse than the other two main clustering techniques.\label{fig:hierarchical}}
\end{figure}

To show the advantage of using the BHV tree space over the Euclidean space, we applied the same clustering methods to Euclidean distance matrices $D(T)$ in $\R^{{n \choose 2}}$, and compared the clustering accuracy obtained. The differences (which are easily noted) are shown in Table \ref{tab.comp_BHV_Eucl}, however we focus only on depth ratios $c=0.8$ and $1.2$ for ease of presentation; the remaining depths also portray the same differences and can be found in the Supplementary Material. We observe that in most cases, the BHV tree space gives better clustering accuracy than when using Euclidean distances. Even though the Euclidean distance of MLE-HKY with KPCA and t-SNE  for $c=0.8$ and direct for $c=1.2$ gives more accurate clustering, these results are much lower than the ones obtained by NJp.

To show the performance of the normalized cut compared to other standard clustering methods in the field, we performed the same experiments using the well-known $k$-means and hierarchical clustering. The k-means clustering is infeasible for BHV tree space, but instead it can be applied with a dimension reduction scheme before putting them to the test. As such, there is no {\em direct} application in the results shown in Fig.~\ref{fig:k-means}. 

From the Figures, we can see that $k$-means is indeed a viable option for accurately clustering the trees, performing similarly to our proposed normalized cut framework. On the other hand, hierarchical clustering has proven to be very bad in clustering in this context, resulting in significantly imbalanced clusters. We tried other linkage methods, but the results were similar.  From our computational study, it can be concluded that normalized cut is very effective in reproducing the cluster structure in gene trees when using BHV distances; moreover, the NJp method is superior in recognizing clusters when compared to MLE methods. Hierarchical clustering, on the other hand, is not recommended in this context. Last, a main advantage of the normalized cut framework is that it requires no dimension reduction, but instead can be directly applied on the BHV space. Computational results of clustering accuracy for $k$-means are also presented in Table \ref{tab.compKMEANS}; the results for hierarchical clustering, as it performs poorly, are omitted but are provided for completeness in the Supplementary Material.

\begin{table}[t]
\centering
{\footnotesize
\begin{tabular}{c|ccc|ccc}
\hline
& \multicolumn{3}{c|}{NJp} & \multicolumn{3}{c}{MLE-GTR} \\
  & KPCA & t-SNE & direct & KPCA & t-SNE & direct \\  \hline
BHV    & 0.838 & 0.824 & 0.836 & 0.522 & 0.674 & 0.620 \\
Euclid & 0.840 & 0.860 & 0.750 & 0.740 & 0.826 & 0.740 \\
\hline
\end{tabular}
(a) $c=0.8$\\
\begin{tabular}{c|ccc|ccc}
\hline
& \multicolumn{3}{c|}{NJp} & \multicolumn{3}{c}{MLE-GTR}  \\
  & KPCA & t-SNE & direct & KPCA & t-SNE & direct \\ \hline
BHV    & 0.988 & 0.976 & 0.982 & 0.872 & 0.944 & 0.738 \\
Euclid & 0.958 & 0.962 & 0.954 & 0.828 & 0.936 & 0.828 \\
\hline
\end{tabular}
(b) $c=1.2$
}
\caption{Comparison of clustering accuracy between BHV space and Euclidean space when using {\em normalized cut}. Euclidean distance gives worse results than geodesic distances in the BHV tree space.  BHV geodesic distance with NJp tree construction is the most suitable for clustering.}\label{tab.comp_BHV_Eucl}
\end{table}

\begin{table}[t]
\centering
{\footnotesize
\begin{tabular}{c|ccc|ccc}
\hline
& \multicolumn{3}{c|}{NJp} & \multicolumn{3}{c}{MLE-GTR} \\
  & KPCA & t-SNE & direct & KPCA & t-SNE & direct \\  \hline
BHV    & 0.838 & 0.806 & N/A & 0.618 & 0.648 & N/A \\
Euclid & 0.842 & 0.840 & 0.830 & 0.740 & 0.826 & 0.740 \\
\hline
\end{tabular}
(a) $c=0.8$\\
\begin{tabular}{c|ccc|ccc}
\hline
& \multicolumn{3}{c|}{NJp} & \multicolumn{3}{c}{MLE-GTR}  \\
  & KPCA & t-SNE & direct & KPCA & t-SNE & direct \\ \hline
BHV    & 0.992 & 0.984 & N/A & 0.742 & 0.866 & N/A \\
Euclid & 0.956 & 0.962 & 0.956 & 0.828 & 0.936 & 0.828 \\
\hline
\end{tabular}
(b) $c=1.2$
}
\caption{Comparison of clustering accuracy between BHV space and Euclidean space when using $k${\em-means}. We observe that the normalized cut framework and $k$-means both perform well. There is no direct application of $k$-means on the BHV space, which is a main advantage of the normalized cut. }\label{tab.compKMEANS}
\end{table}

\subsection{Genome data set on coelacanths, lungfishes, and tetrapod}\label{subsec:lungfish}

On top of the simulated datasets, we have also applied the clustering methods to the dataset comprising 1,290 nuclear genes encoding 690,838 amino acid residues obtained from genome and transcriptome data by \cite{Liang2013}. Over the last decades, the phylogenetic relations between coelacanths, lungfishes, and tetrapods have been controversial despite the existence of many studies on them \cite[]{Hedges2009}. Most morphological and paleontological studies support the hypothesis that the lungfishes are closer to the tetrapods than coelacanths (Tree 1 in Figure 1 from \citet{Liang2013}), however, there exists research in the field that supports the hypothesis that the coelacanths are closer to the tetrapods  (Tree 2 in Figure 1 from \citet{Liang2013}).  Others support the hypothesis that the coelacanths and the lungfishes form a sister clades (Tree 3 in Figure 1 from \citet{Liang2013}) or tetrapodes, lungfishes, and coelacanths cannot be resolved (Tree 4 in Figure 1 from \citet{Liang2013}). In this subsection, we apply the normalized cut framework for clustering to the genome data set from \cite{Liang2013} and analyze each obtained cluster.

\begin{figure*}[t]
  \centering
  \includegraphics[width=6.5cm]{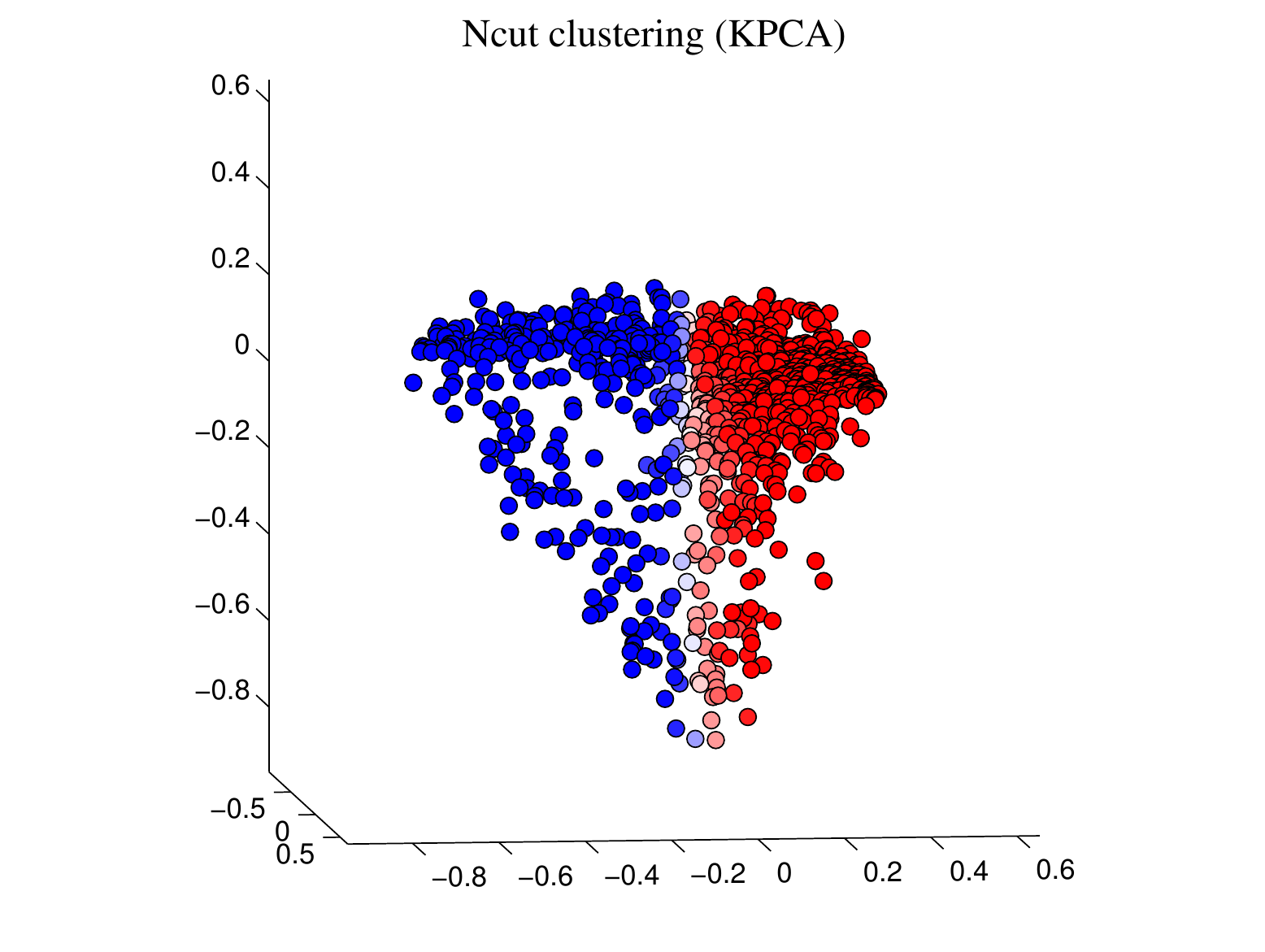}
  \includegraphics[width=6.5cm]{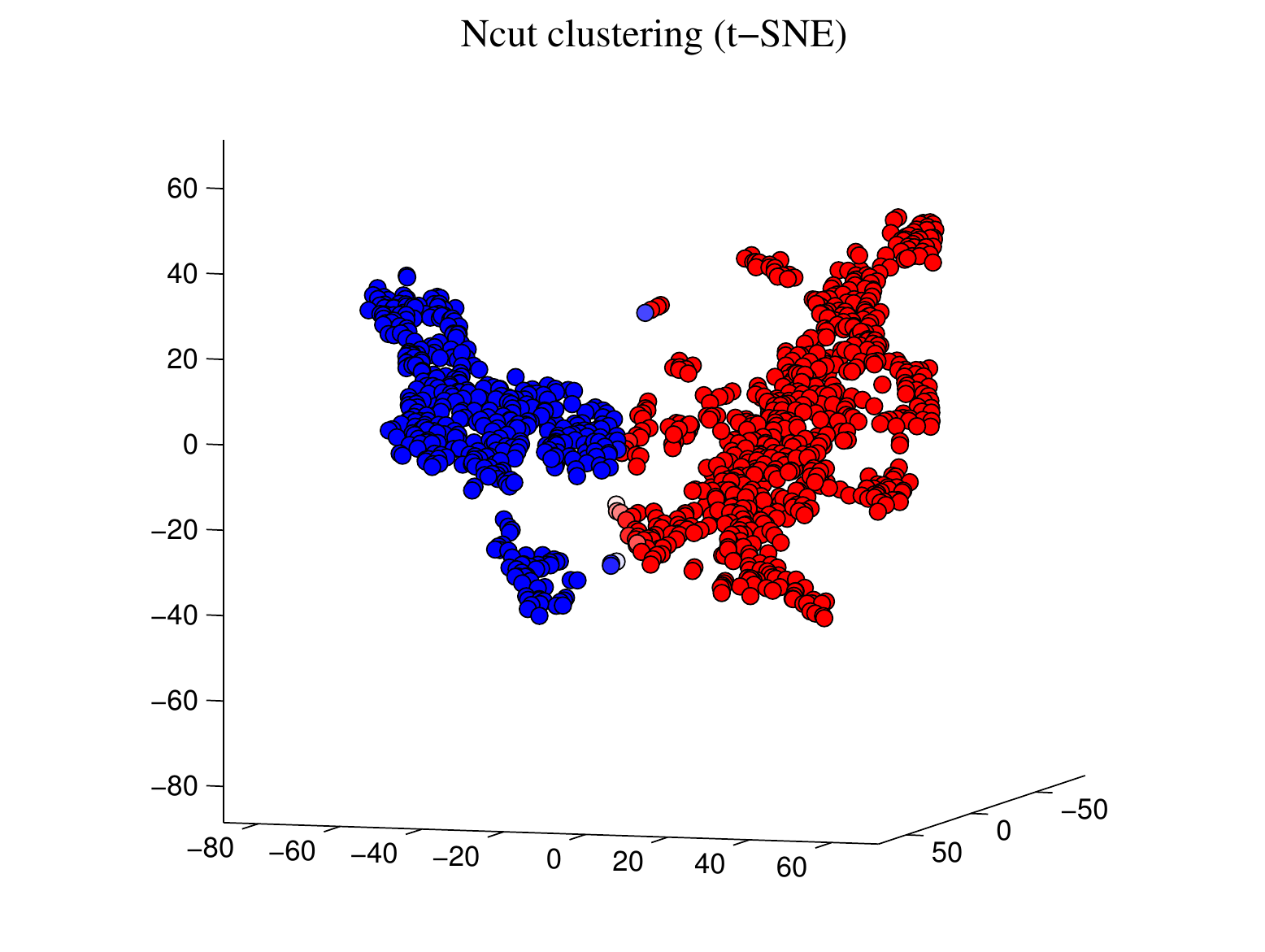}\\
  \caption{Clustering of the genome data set.  The two clusters are depicted in red and blue with bootstrap confidence shown by color density.}\label{fig:fish_clustering}
\end{figure*}

We applied the clustering methods (with and without a dimension reduction) to the distance matrix computed from the set of gene trees constructed by the NJp method. The number of clusters in Ncut was set to two, that is, a bipartition. Fig.~\ref{fig:fish_clustering} shows the clustering results with KPCA and t-SNE, plotted on the three dimensional space found by the dimension reduction.  The red and blue colors show the two clusters, where the color density represents the bootstrap confidence explained below.

To evaluate the stability of clustering, we computed a bootstrap confidence probability for each gene.  Namely, given $N\times N$ distance matrix $(D_{ij})$ as an input to the Ncut framework, we generated random resampling $\{i_1,\ldots,i_N\}$ from $\{1,\ldots,N\}$ with replacement, and applied Ncut to $(D_{i_a i_b})_{a,b=1}^N$.  We repeated this procedure 100 times with independent random indices, and computed the ratio that a gene is classified in the same cluster as the one given by $(D_{ij})$.

\begin{figure*}[t]
  \centering

  \includegraphics[width=5.0cm]{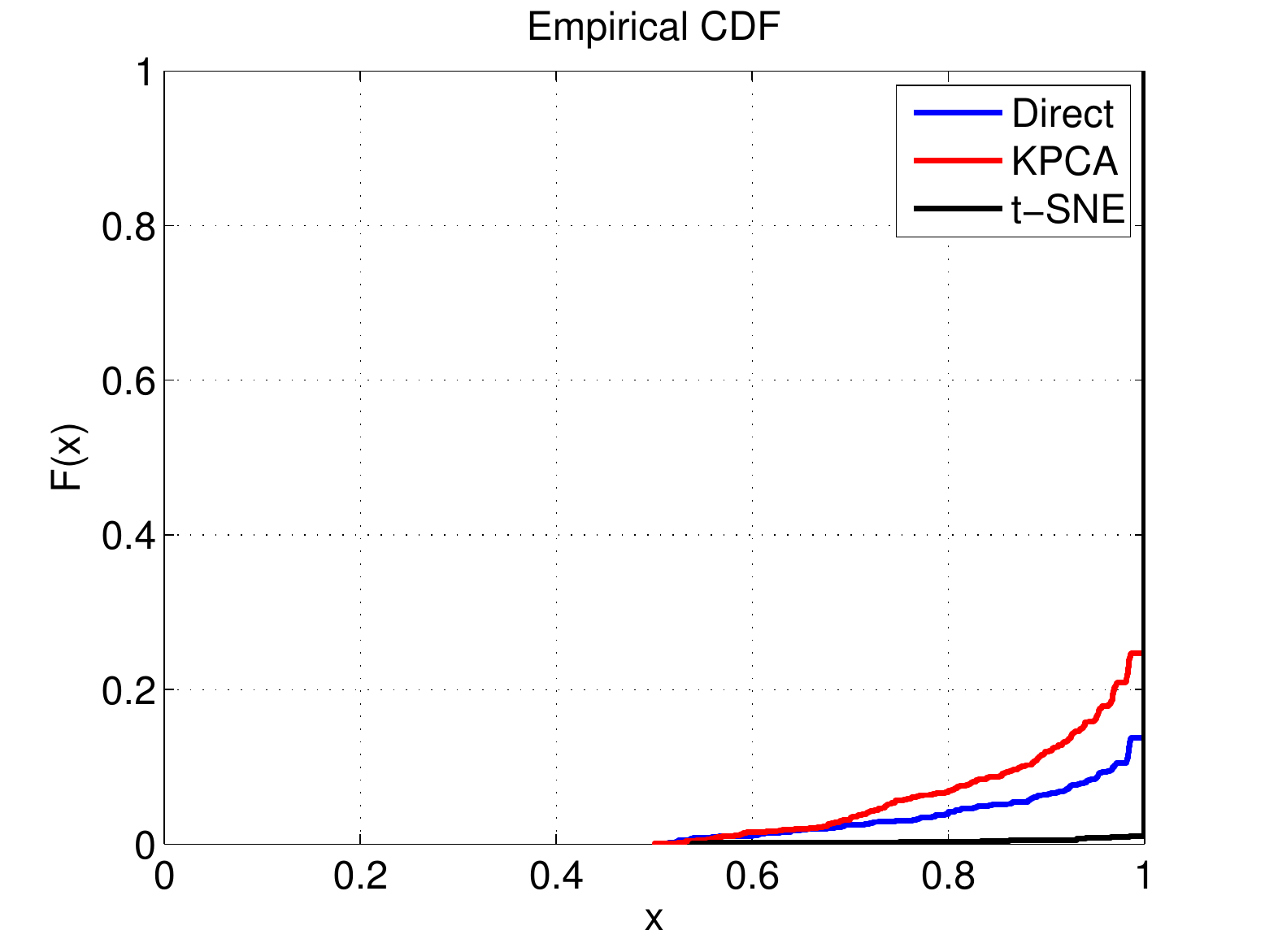}
  \includegraphics[width=5.0cm]{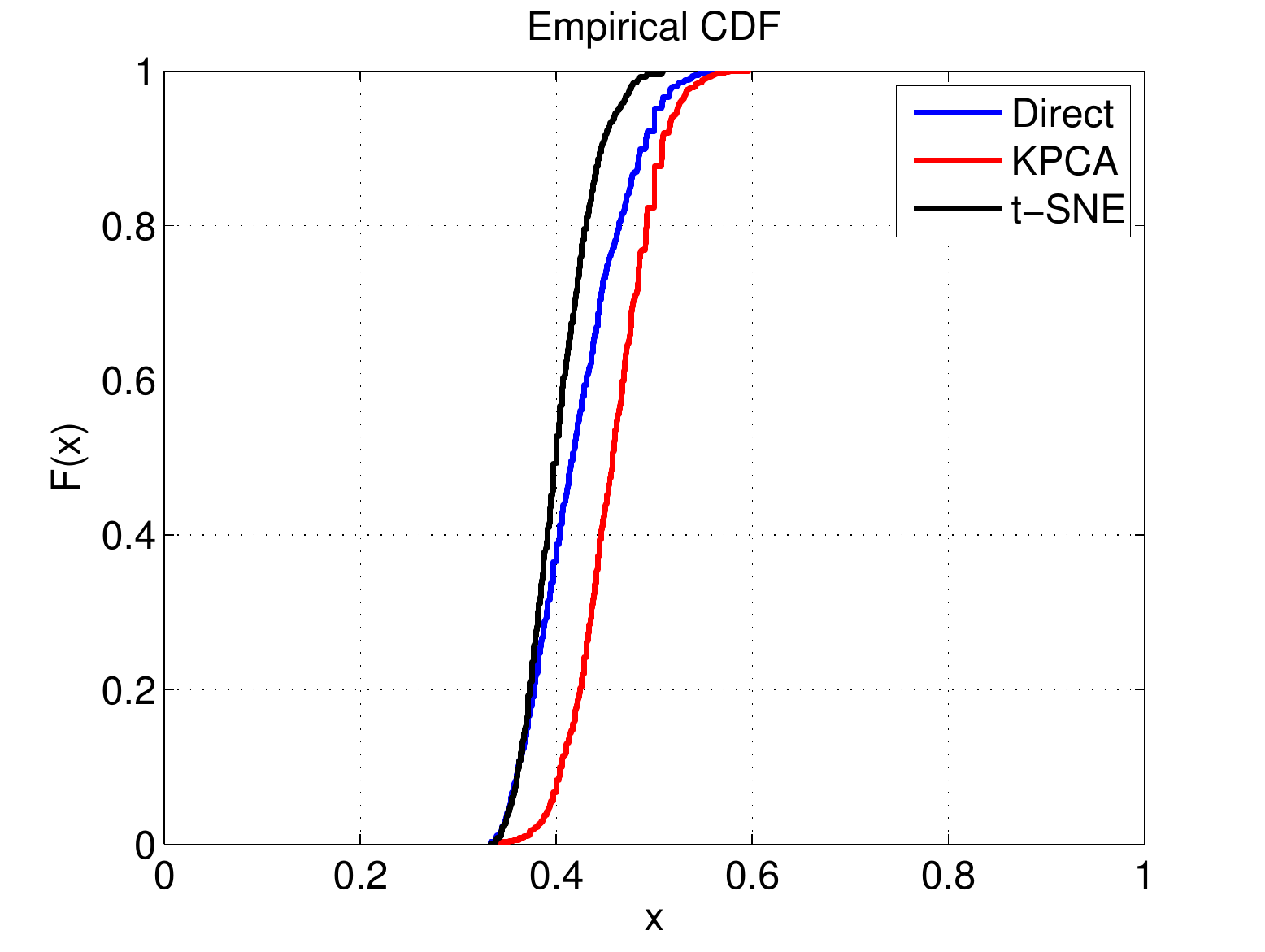}
  \caption{Cumulative distribution functions of confidence values for clustering. The two clusters (left) are reliable, while the three clusters (right) are unstable.}\label{fig:cdf_clustering}
\end{figure*}

We computed the bootstrap confidence for all 1,290 genes.  The cumulative distribution functions of these values are shown for the tree clustering methods in Fig.~\ref{fig:cdf_clustering} (left).  The ratio of genes with confidence above 0.95 is 91.4\%, 83.8\%, and 99.2\% for direct Ncut, KPCA Ncut, and t-SNE Ncut, respectively.  For comparison, we computed the bootstrap confidence for Ncut with three clusters. Fig.~\ref{fig:cdf_clustering} (right) shows the cumulative distribution function, which clearly reveals that three clusters are unstable.  From these observations, we see that the two clusters obtained by the methods are not artifacts but a stable structure in the genome data.

The clusters obtained by the three methods look different in their shapes. We then examined agreements of the clusters at the gene level.  After extracting the genes with bootstrap confidence not less than TH, (TH $= 0.90$ or $0.95$), we evaluated the agreement of methods A and B by
\[
    t_{AB}:=\frac{|C^1_A \cap C^1_B| + |C^2_A\cap C^2_B|}{N_A},
\]
where $N_A$ is the number of genes by Method A with confidence larger than TH and $C^i_A$ is  the $i$-th ($i=1,2$) cluster by Method A ($N_A=|C_A^1|+|C_A^2|$).  We identified which cluster in A corresponds to a cluster B by the number of common genes.  Table \ref{tbl:agree} shows the value $t_{AB}$ for every pair of the three methods.  We can see that majority of genes in a cluster agrees to another cluster given by a different method.  This confirms that the clustering reveals the structure of the data.  KPCA Ncut and t-SNE Ncut are slightly less consistent, which may be caused by the difference of $N_A$ for the two methods.

\begin{table}
\centering
{\footnotesize
(a) TH=0.90\\
  \begin{tabular}{c|ccc|c}
  $A\backslash B$ & direct &  KPCA & t-SNE & $N_A$\\
  \hline
    Direct &  - &    0.917 &   0.800 & 1207 \\
    KPCA &    0.912  & -  &    0.757  & 1135\\
    t-SNE &    0.812  &  0.785 &  - & 1284 \\
    \hline
\end{tabular}

(b) TH = 0.95\\
  \begin{tabular}{c|ccc|c}
  $A\backslash B$ & direct &  KPCA & t-SNE & $N_A$\\
  \hline
    Direct & - &    0.896  &  0.786 &  1180\\
    KPCA &    0.886  & - &   0.712 & 1081\\
    t-SNE &    0.803  &  0.757 &  - & 1280\\
    \hline
\end{tabular}
}
  \caption{Agreement of clusters among the three methods for the normalized cut. The rightmost column shows the number of selected genes for each method ($N_A$). }
  \label{tbl:agree}
\end{table}

Finally we conducted the phylogenetic analysis on the clusters of gene trees. For each clustering method (direct Ncut, KPCA Ncut, and t-SNE Ncut), we have reconstructed a consensus tree from each cluster.  To construct the consensus tree, we have used the gene trees in each cluster with bootstrap value greater than $0.95$ and took the majority rule with more than $50\%$ for reconstructing the consensus tree for resolving each split on the tree. With all the clustering methods, the result suggests that there are two clusters in the genome-wide data set on coelacanths, lungfishes, and tetrapods: the number of genes are $(858, 322)$, $(761, 320)$, and $(817, 463)$ for direct Ncut, KPCA Ncut and t-SNE Ncut, respectively.  Note that we used only gene trees with bootstrap confidence $\geq 0.95$. 

\begin{figure*}[b]
  \centering
\includegraphics[width=4cm]{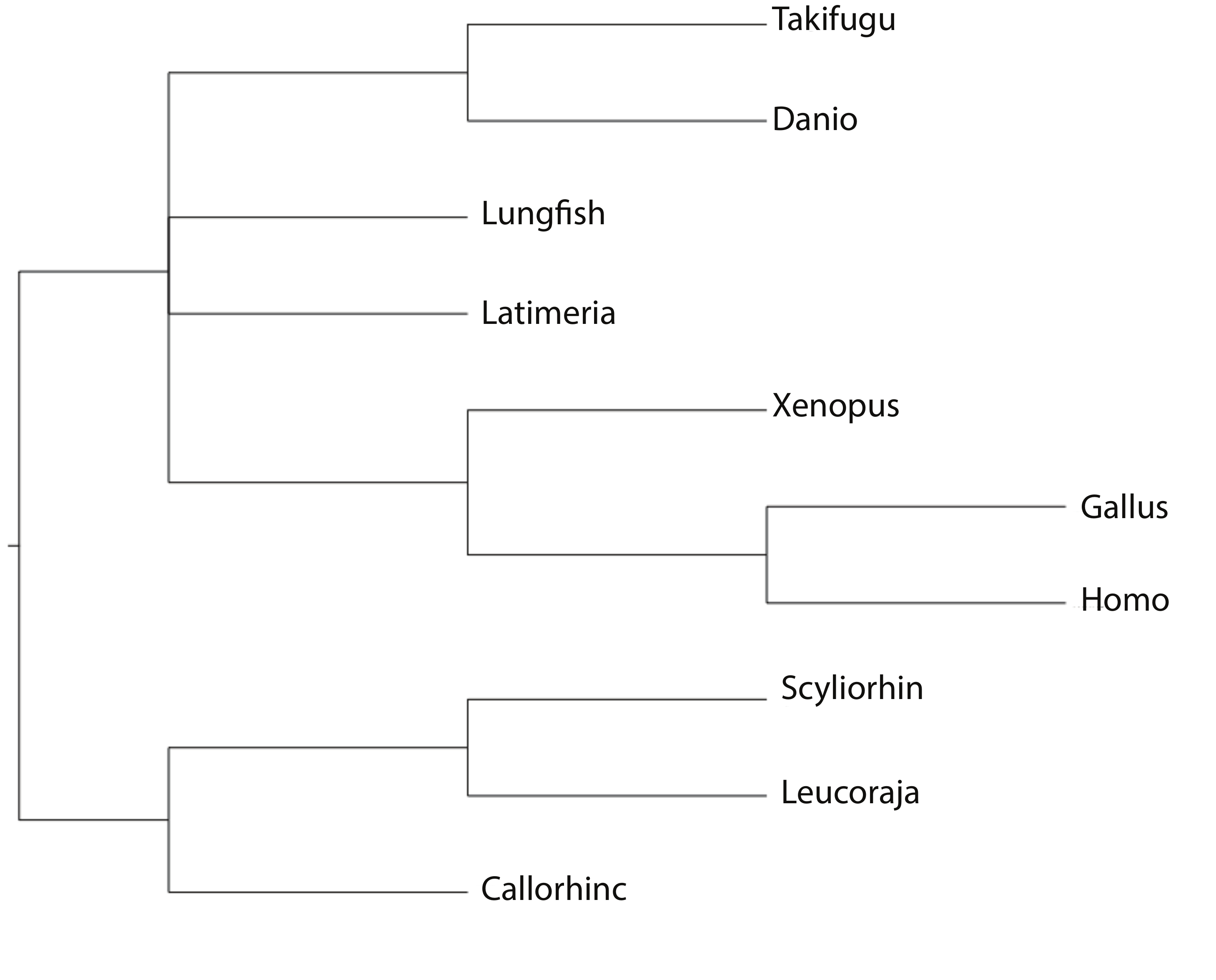}
   \includegraphics[width=4cm]{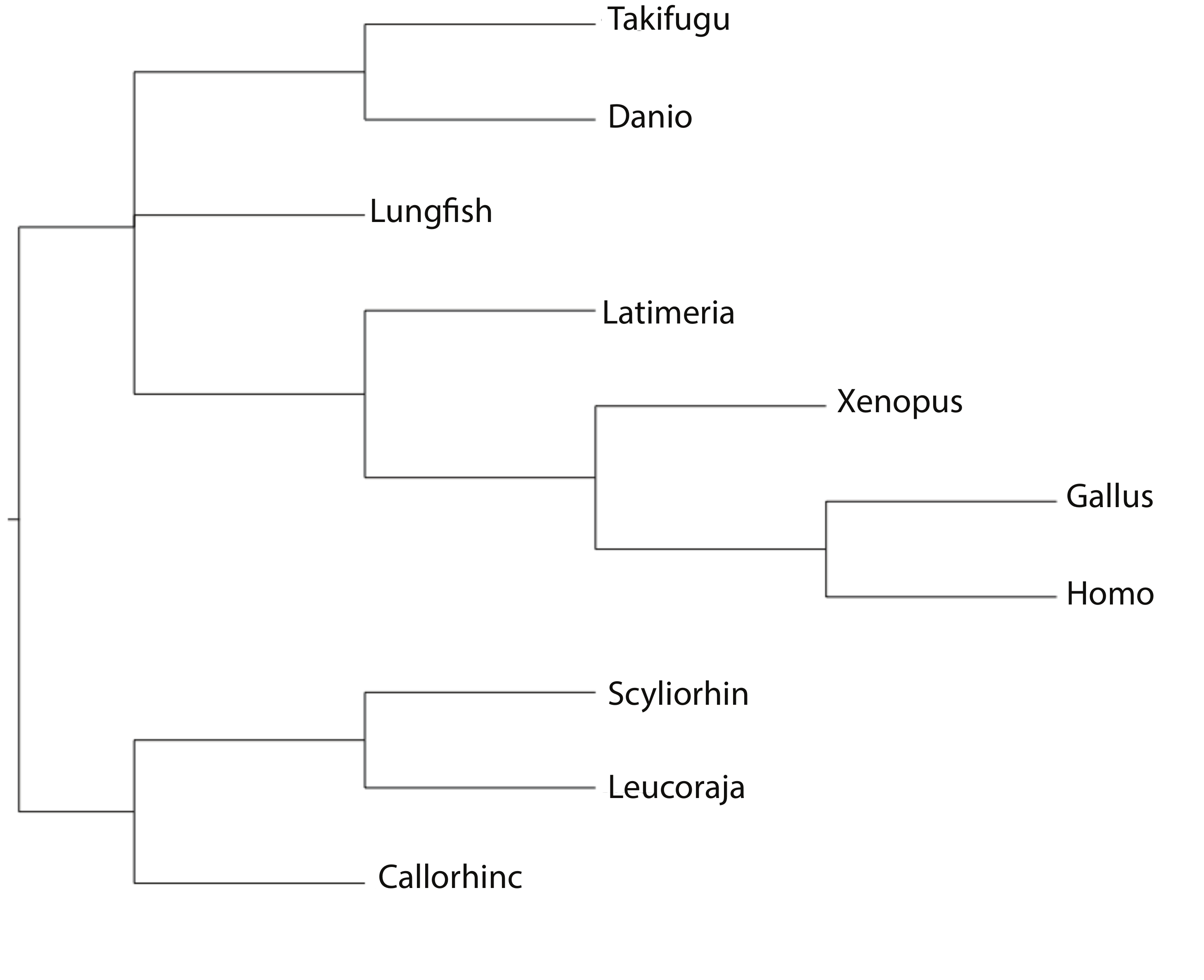}
  \caption{The majority rule consensus tree consists of gene trees
    with more than $0.95$ bootstrap values in each cluster.  Each
    split in the trees is resolved only if we have majority, i.e. 50\%
    of all given gene trees in each set agree.}\label{fig:con_tree}
\end{figure*}

With all of the three methods, direct Ncut, Ncut with KPCA, and Ncut with t-SNE, one cluster of the gene trees provides the tree topology Tree 4 from Figure 1 in \citet{Liang2013}, while the other cluster gives the tree topology Tree 2 from Figure 1 in \citet{Liang2013} (see Fig.~\ref{fig:con_tree}).

\begin{figure*}[b]
  \centering
  \includegraphics[width=4cm]{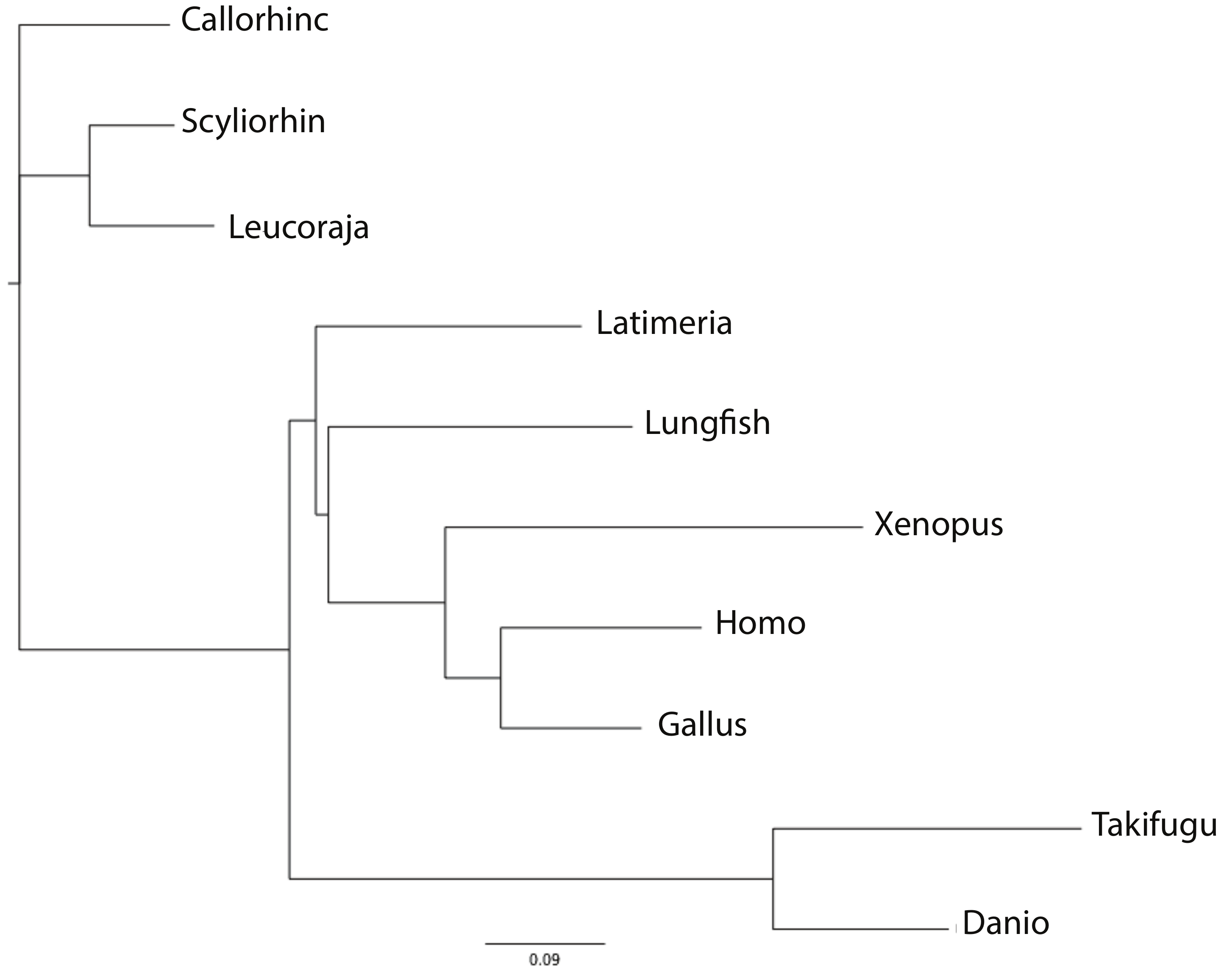}
  \includegraphics[width=4cm]{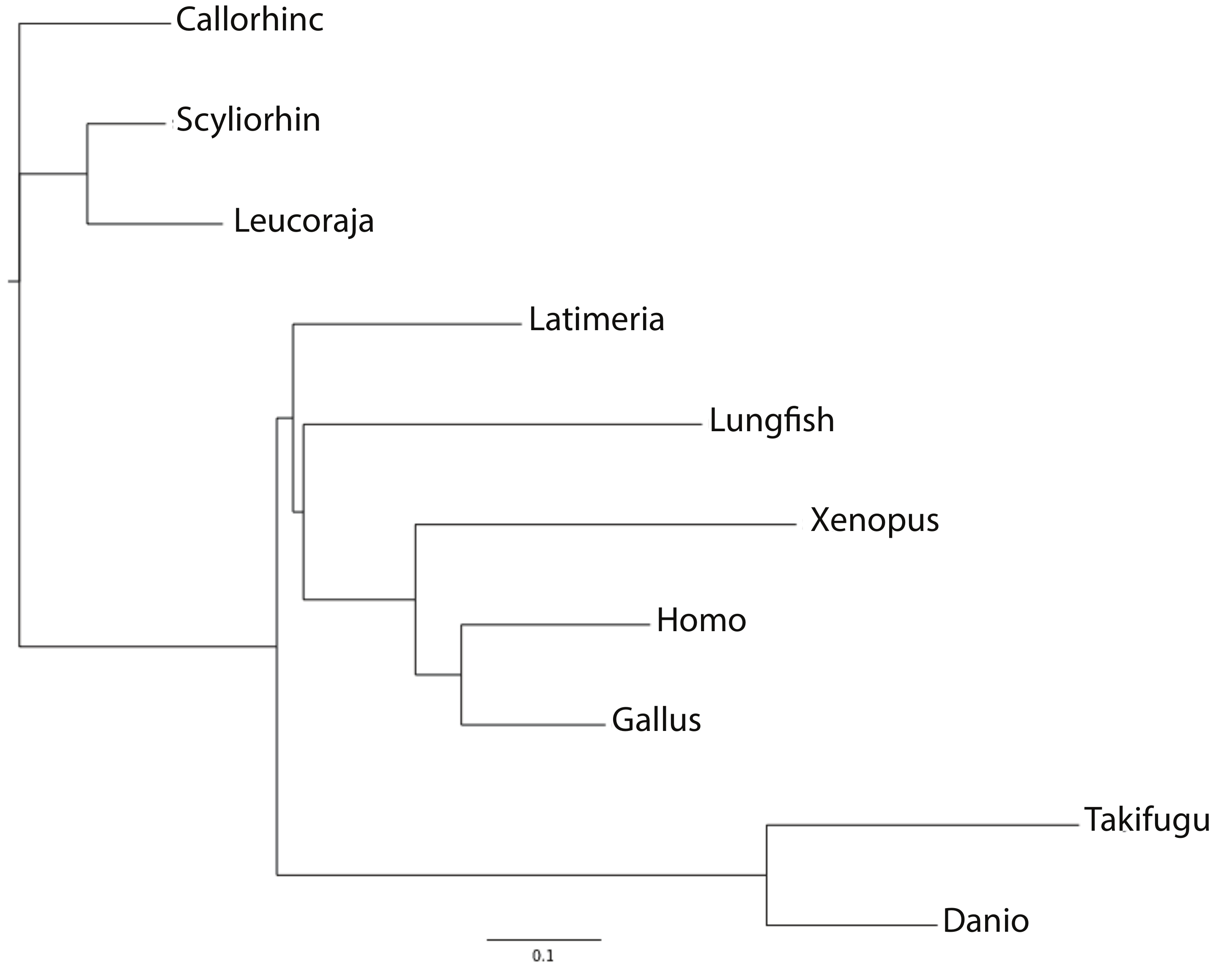}
  \vskip 3mm
  \caption{The reconstructed trees obtained by concatenating the alignments from each cluster after using direct Ncut. For this result, we employed the Bayesian inference using the software {\tt PhyloBayes 3.3} under a mixture model CAT $+ \Gamma 4$ with two independent MCMC runs for 10,000 cycles.}\label{fig:PB_tree}
\end{figure*}

We have also reconstructed a tree from each cluster by concatenating the alignments using the software {\tt PhyloBayes 3.3} under a mixture model CAT $+ \Gamma 4$ with two independent MCMC runs for 10,000 cycles. However, we did not observe any difference in the tree topologies, i.e., the reconstructed trees have all the same tree topology as Tree 1 in from Figure 1 in \citet{Liang2013} (see Fig.~\ref{fig:PB_tree}).

\section{Discussion}
\label{discussion}

In this paper we have shown three main results: the Ncut clustering algorithm works well on the gene trees reconstructed via the NJp under the evolutionary models; via the Ncut clustering algorithm we found two clusters on the genome data sets from \cite{Liang2013}; last, $k$-means performs equally well after dimension reduction, while hierarchical clustering is always outperformed in this context. 

\noindent
{\em Simulations}:
As we have shown by simulations in Section \ref{results}, the normalized cut framework works effectively on the set of gene trees reconstructed via the NJp method compared to the trees reconstructed via the MLE under the evolutionary models (Table  \ref{tbl:agree}, Fig. ~\ref{fig:cluster_acc} as well as Fig. ~\ref{fig:k-means} and \ref{fig:hierarchical} for the other methods).  It is not clear why this phenomenon appears in our computational study and it is of interest to investigate mathematically the reason it happens.

\noindent
{\em Coelacanths, lungfishes, and tetrapods data set}:
Using the Ncut algorithm on the gene trees reconstructed via the NJp method, we were able to identify two clusters.  Bootstrap confidence analysis suggests that these are two reliable clusters and it appears to be very unlikely to have more than two clusters (see Fig. \ref{fig:cdf_clustering}). From the two clusters we were able to find using the Ncut framework, we have reconstructed the consensus trees and their tree topologies did not support the hypothesis that the lungfishes are the closest living relatives of the tetrapods as in \cite{Liang2013}, but supported the hypotheses that the coelacanths are most closely related to the tetrapods, and that the coelacanth, lungfish, and tetrapod lineages diverged within a very short time interval.  Since clustering analysis with Ncut does not infer any evolutionary events that caused the clusters, it would be interesting and important to further investigate how these clusters were made in the evolutionary history.

\section*{Funding}

This work has been supported by JSPS KAKENHI 26540016. \vspace*{-12pt}

\bibliographystyle{natbib}
%
%
\bibliography{samsi}

\end{document}